\documentclass[12pt]{iopart}
\usepackage{graphicx}
\usepackage{subfigure}
\usepackage{cite}

\begin{document}

\title[Muon Arrival Time Distributions]{Features of Muon 
Arrival Time Distributions of High Energy EAS at Large
Distances From the Shower Axis}

\author{I. M. Brancus\dag,
H. Rebel\ddag \S , A. F. Badea$\|$
\footnote[7]{on leave of absence from National Institute of 
Physics and Nuclear Engineering, 7690 Bucharest, Romania}, 
A. Haungs\ddag,\\ 
C. D. Aiftimiei\dag, J. Oehlschl{\"a}ger\ddag $\ $ and M. Duma\dag}

\address{\dag\ National Institute of Physics and Nuclear 
Engineering, 7690 Bucharest, Romania}
\address{\ddag\ Forschungszentrum Karlsruhe, Institut f{\"u}r 
Kernphysik, 76021 Karlsruhe, Germany}
\address{\S\ University of Heidelberg, Faculty of Physics and 
Astronomy, 69133 Heidelberg, Germany}
\address{$\|$\ University of Karlsruhe, Institut f{\"u}r 
Experimentelle Kernphysik, 76021 Karlsruhe, Germany}

\begin{abstract}

In view of the current efforts to extend the KASCADE experiment 
(KASCADE-Grande) for observations of Extensive Air
Showers (EAS) of primary energies up to 1$\cdot 10^{18}$ eV, the 
features of muon arrival time distributions and their correlations 
with other observable EAS quantities have been scrutinised on 
basis of high-energy EAS, simulated with the
Monte Carlo code CORSIKA and using in general the QGSJET model as 
generator. Methodically various correlations of adequately
defined arrival time  parameters with other EAS parameters 
have been investigated by invoking non-parametric methods for
the analysis of multivariate distributions, studying the 
classification and misclassification probabilities of various
observable sets. It turns out that adding the arrival time 
information and the multiplicity of muons spanning the
observed time distributions has distinct effects improving the mass 
discrimination. A further outcome of the studies is the feature that 
for the considered ranges of primary energies and of distances 
from the shower axis the discrimination power of global arrival time 
distributions referring to the arrival time of the shower 
core is only marginally enhanced as compared to local distributions 
referring to the arrival of the locally first muon.

\end{abstract}

\pacs{96.40 Pq; \\
Keywords: Cosmic rays; Airshowers; 
Muon arrival time distributions.}


\maketitle

\newpage





\section{Introduction}

Since the first experimental studies in 1953 by Bassi, Clark and 
Rossi \cite{01} and Jelley and Whitehouse \cite{02} arrival time 
distributions of the charge particle components of Extensive Air 
Showers (EAS) have been often theoretically and experimentally 
investigated under various aspects \cite{03,04,05,06,07,08,09}. 
Systematic studies are performed particularly at large EAS
detector installations: in Haverah Park \cite{10,11,12,13,14,15,16}, 
at the Potchefstroom University \cite{17}, in Akeno and on Mt. 
Chacaltaya \cite{18,19,20}, at MSU \cite{21}, more recently with 
the GREX/ COVER-PLASTEX \cite{22,23,24,25,26} and
KASCADE \cite{27,28,29,30,31,32,33,34}
experiments. Measurements of arrival time distributions of high energy 
hadrons near the shower core have been
reported by the Maryland University group \cite{35} and by KASCADE 
\cite{36}. In addition to the fact that the observed arrival time 
distributions of the EAS particles display the phenomenological 
appearance (shape and structure) of the EAS disc, they provide also
a coded picture of the longitudinal EAS development. Especially the 
muon component, when multiple scattering in the
atmosphere at higher muon energies gets negligible, maps the 
distributions of the production heights via the time-of-flight
of the muons from the loci of decay of the charged parent pions to 
the detector \cite{27,37}. Fig.~1 sketches the relation in a
simplified way. Muons released in higher atmospheric altitudes (and 
observed on ground at  fixed larger distances from the
shower axis) show smaller delays relative to the arrival of the 
shower centre ("light front"), and their (relative)
arrival times let expect some discrimination features for the mass of
the EAS primary \cite{05,06,07,27,29}. 
Thus the arrival time distributions of muons, produced in iron induced 
showers e.g. should get shifted to shorter delays due to the faster
development as compared to proton induced showers of the same primary 
energy. Of course, a serious analysis of this
mapping has to invoke realistic simulations of the EAS development and 
of the muon tracking, taking into account the
influence of multiple scattering and off-axis production \cite{37} 
of the muons. Fig. 1 indicates also that observations of
the angle distributions of the muon incidence provide alternative 
experimental possibilities \cite{38,39}, in case that
multiple scattering effects do not seriously obscure that information 
\cite{37}. The information content of the combination
of  both types of EAS observations, known as 
"Time-Track-Complimentary-Principle" \cite{37,40}, 
has been scrutinised in Ref. \cite{29}.

\begin{figure}[!t]
\centering
\includegraphics[bb=187 202 492 675, angle=270, width=15.0cm]{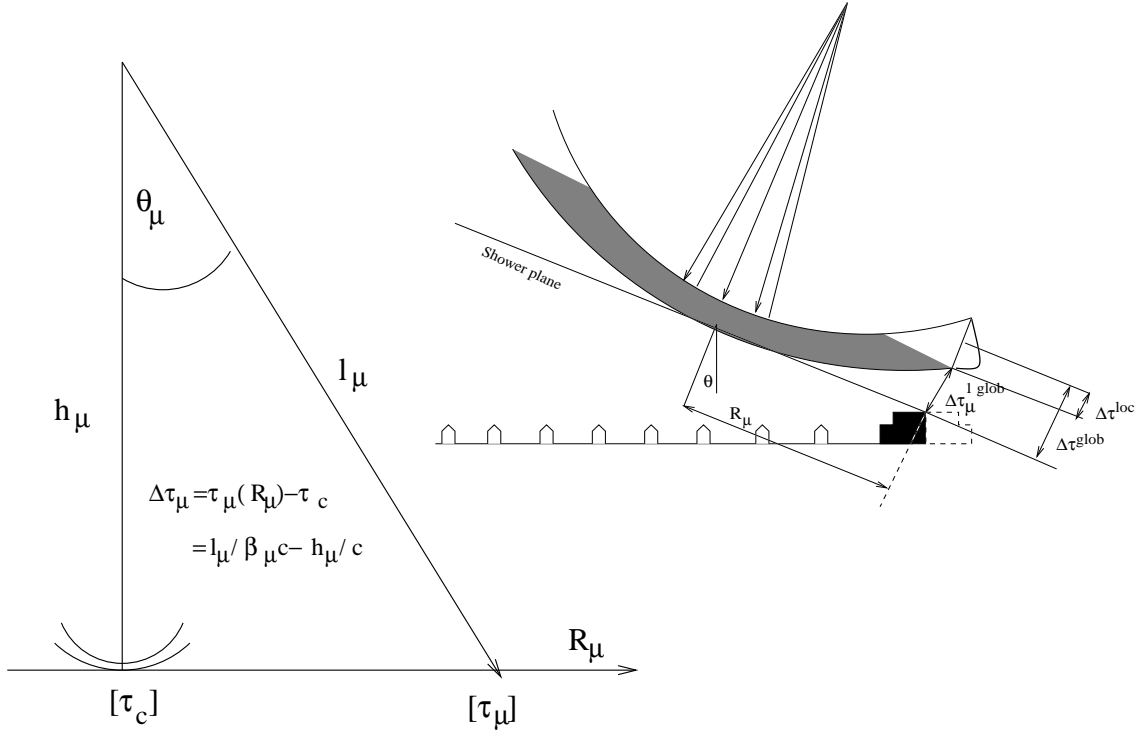}
\caption{\label{Fig. 1} Sketch of a simplified relation between muon 
arrival time and angle of muon incidence, respectively,
with the production height, neglecting multiple scattering and 
off-axis production ($\beta _{\mu} c$ = velocity of the muon). 
Additionally (right) the definitions of the arrival time quantities 
characterising the shape and the structure of the muon disc are indicated.}
\end{figure}

Arrival times $\tau ^{1}_\mu < \tau ^{2}_\mu < \tau ^{3}_\mu$,... of the 
EAS muons, locally registered by timing detectors at a distance $R _\mu$
from the shower axis have to refer to a well defined zero-time, usually 
the arrival time $\tau _{c}$ of the shower core (global arrival times):

\begin{equation*}
\Delta \tau ^{n\ glob} _\mu = \tau ^n _\mu (R_\mu) - \tau _c
\end{equation*}

e.g.
\begin{equation*}
\Delta \tau ^{1\ glob}_\mu = \tau ^1 _\mu (R_\mu) - \tau _c
\end{equation*}

Often there are experimental difficulties  to determine the arrival 
time $\tau _c$ with sufficient experimental precision, and
only "local" times are considered. They refer to the foremost 
(first locally registered) muon:

\begin{equation*}
\Delta \tau ^{n\ loc} _\mu (R_\mu) = \tau ^n _\mu (R_\mu) - \tau ^1 _\mu (R_\mu)
\end{equation*}

{\it Local} arrival time distributions display the internal 
structure of the shower disc (characterised by various time
parameters), but they do not carry information about the 
shape (curvature) of the front. Nevertheless studies based on
simulations of the EAS development \cite{16,19} have shown that 
mass discrimination effects are just pronounced in observations
of $ \Delta \tau ^{1\ glob} _\mu $.

For event-by-event observations with a fluctuating number of 
muons, the single relative arrival time distributions can
be characterised by the mean values $\Delta \tau _{mean}$, and by 
various quantiles $\Delta \tau _q$, like the median
$\Delta \tau _{0.50}$, the first quartile $\Delta \tau _{0.25}$
and the third quartile $\Delta \tau _{0.75}$.
For ordered statistics of measured times
$\Delta \tau _1 \le \Delta \tau _2 \le . . . \le \Delta \tau _n$
and $k := n \alpha + \xi, k$ integer and $\xi \in [0,1)$, the
$\alpha$-quantile $\Delta \tau _\alpha$ ~ $(\alpha \in (0,1))$ is the 
following \cite{28,29}:
\begin{equation*}
\Delta \tau _\alpha = \left\{ \begin{array}{ll}
( \Delta \tau _k + \Delta \tau _{k+1} ) / 2 : & for \: \: \xi = 0 \\
  \Delta \tau _k : & for \: \: \xi \in (0,1) \end{array} \right. 
\end{equation*}
That means: in the case of large $n$, a fraction $\alpha$ 
of muons has arrival times less than $\Delta \tau _\alpha$. The 
{\it mean values} and {\it dispersions} (standard deviations) represent the
time profile of the EAS muon component. Measurements of muon arrival 
time distributions and the determination of the
distributions of various different time quantities have been a 
subject of the current investigations of the KASCADE
experiment \cite{27,28,30,31,32,33,34}. In these investigations 
the temporal EAS structure has been studied in detail in dependence on the
shower size $N_e$, the muon number $N_\mu$, or the truncated muon 
number $N _\mu ^{tr}$ (which is used as approximate energy identifier in
KASCADE observations \cite{41}), respectively, and from the angle 
of EAS incidence $\theta$. Special features arising from the observed
muon multiplicity have been revealed \cite{33}. More recently 
\cite{34} using larger samples of shower events, experimentally
observed with KASCADE,  the sensitivity of local muon arrival 
time distributions and of their correlations with other EAS
observable to the mass of the EAS primaries has been investigated. 
Methodically advanced statistical analysing techniques
have been applied \cite{42}, based on Bayesian decision making 
for the classification of the EAS events. The procedure requires
simulated distributions (processed through the detector response) 
as reference patterns, provided by the Monte Carlo
simulation program CORSIKA \cite{43} with a particular model of 
the hadronic interactions. Thus the analysis necessarily implies
some model dependence. But the consistency with the invoked QGSJET 
model \cite{44} could be established thanks to the
measurements with varying distance $R_\mu$ from the shower centre  
and varying muon multiplicity thresholds for being accepted
in the observation sample. Nevertheless, it has been also shown 
that the mass discrimination power of local muon arrival
time distributions is rather marginal in the observed range 
$R_\mu < 100$ m and primary energies $E_0 \le 10^{16}$ eV, and 
it does not significantly help for the classification.

The phenomenological features of the time structure of high energy 
EAS at larger distances from the shower axis have
been experimentally studied with the Haverah Park 
\cite{10,11,12,13,14,15,16} and the Akeno air shower arrays 
\cite{18,19,20}, and the results have
been compared with phenomenological model predictions 
(scaling models and multiplicity prescriptions for the particle
production in hadronic collisions). An analysis of the mass 
discrimination power on basis of modern models of the hadronic
interactions has not yet been performed for these cases.

The recently started extension of the KASCADE detector installation: 
KASCADE-Grande \cite{45} will not only provide the
possibility to extend the studies of muon arrival time distributions 
and their correlations to larger distances $R_\mu$ and
larger energies of the observed EAS. There is also a chance to 
measure global arrival times, at least for $R_\mu$ up to
350 m \cite{46}. The investigations of this paper use the analysis 
techniques of our previous studies \cite{29,34} and refer mainly
to the QGSJET \cite{44}, in some few comparisons also to other 
models. The studies are focused to explore the
basic information content of muon arrival time distributions of EAS 
with energies  up to ca. 1$\cdot 10^{18}$ eV. The analyses are
completely based on data of simulated showers and consider ideal 
cases, i.e. neglecting the influence of the detection
system. The view of interest is the discrimination of the mass of 
the primary particles (with adopting a particular
hadronic model as generator of the reference patterns).
For some observable combinations the differences in the predictions of
Monte Carlo simulations using different high-energy models as generators
have been scrutinised. Only marginal differences for the arrival time
distributions have been found. 

\section{Time profiles of the EAS muon component}

The present studies consider predictions of realistic and detailed 
Monte Carlo simulations of the EAS development
initiated by cosmic ray primaries of different mass in the primary 
energy range up to 1$\cdot 10^{18}$ eV. The simulations have
been performed  by use of the program CORSIKA (version 5.64) \cite{43}, invoking 
in general different models for the hadronic interaction:
QGSJET (version 98) \cite{44}, VENUS (version 4.125) \cite{47} and SIBYLL 
(version 1.6) \cite{48}. For the 
particle interactions below $E_{lab}= 80$ GeV the
GHEISHA \cite{49} option is used. Earth magnetic field, observation 
level and particle detection thresholds have been chosen in accordance 
with the observation conditions of the KASCADE experiment, but without 
a detailed account for the detector response. The U.S.
standard atmosphere (see Ref. \cite{43}) has been adopted for the 
simulations. For a realistic description of the
electron-photon component, instead of the NKG approximation 
\cite{50} the EGS option \cite{51} has been preferred.

\begin{figure}[!b]
\centering
\includegraphics[bb=33 181 564 685,width=13.0cm]{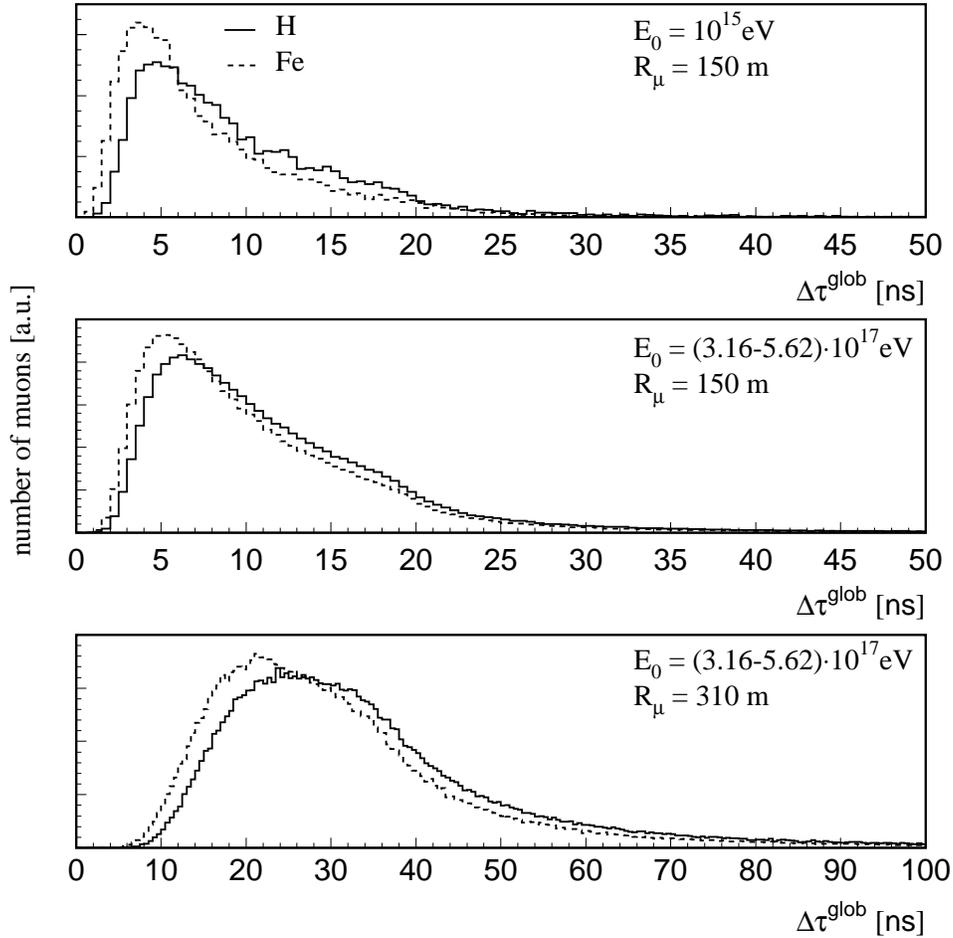}
\caption{\label{Fig. 2} Average global muon arrival time distributions 
to be observed at the distances $R_\mu$ to the
shower core of proton and iron induced EAS at two different 
energies $E_0$ = $10^{15}$ eV (vertical incidence) and 
$E_0$ = (3.16-5.62)$\cdot 10^{17}$ eV ($0^{\circ}-30^{\circ}$)
and with the muon energy detection threshold $E_{thr}$ =240 MeV
(according to the threshold of the KASCADE detector array \cite{31}).}
\end{figure}

\begin{figure}[t]
\centering
  \subfigure{
  \includegraphics[bb=27 153 562 677,width=13cm]{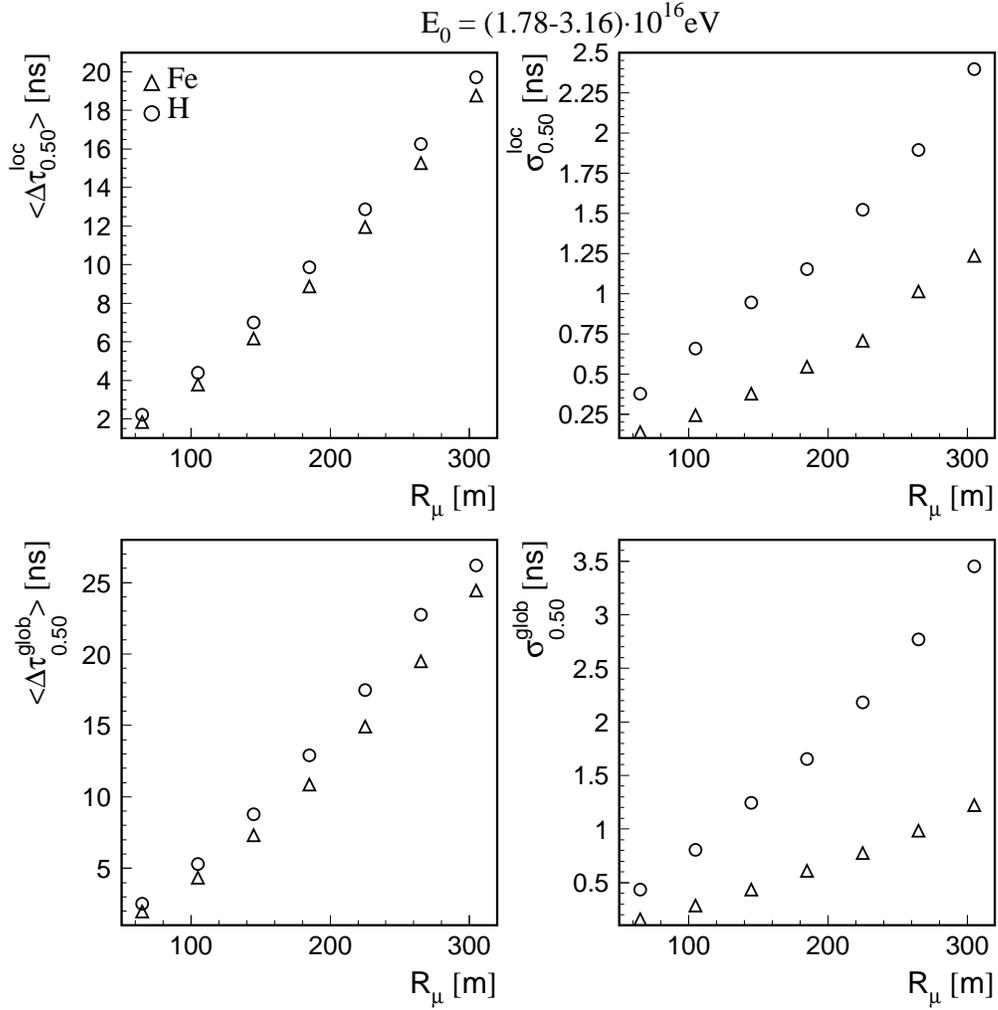}}
\caption{\label{Fig. 3a}{\bf($a$)} Comparison of local and global 
time profiles $<~\Delta \tau _{0.50}~>$ of the EAS muon component,
induced by proton
and Fe primaries of the energy range of (1.78-3.16)$\cdot 10^{16}$ eV,
based on EAS Monte Carlo simulations using CORSIKA and the QGSJET model 
as generator.}
\end{figure}

\addtocounter{figure}{-1}
\begin{figure}[t]
\addtocounter{subfigure}{1}
\centering
  \subfigure{
  \includegraphics[bb=27 153 562 671,width=13.0cm]{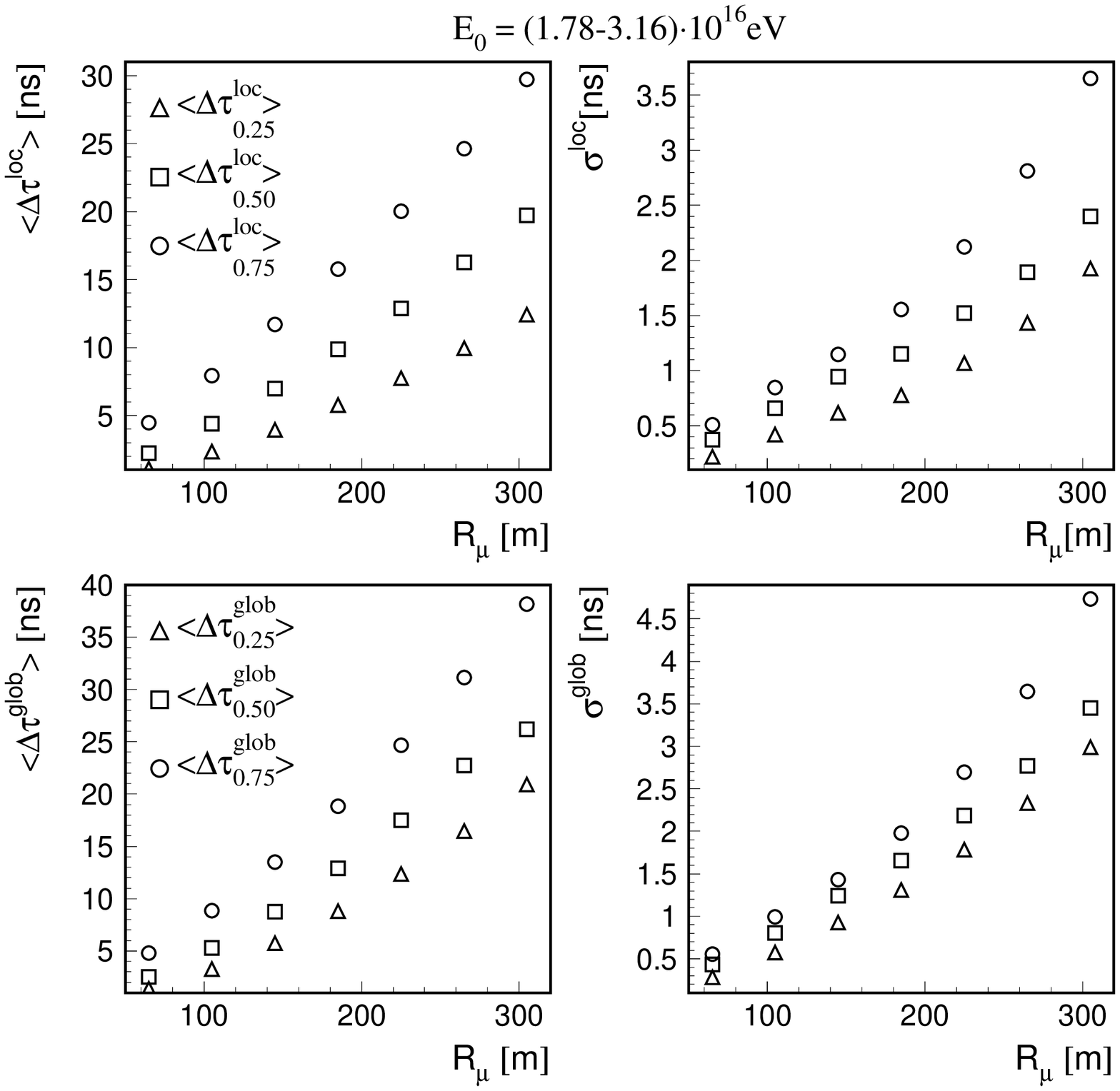}}
\caption{\label{Fig. 3b}{\bf($b$)} Local and global time profiles 
of different quartiles $<\Delta \tau _q>$ of the EAS muon component, 
induced by protons of the energy $E_0$ = (1.78-3.16)$\cdot 10^{16}$ eV
from EAS Monte Carlo simulations using CORSIKA and the QGSJET model
as generator.}
\end{figure}

A first set (I) of simulations comprises samples (of ca. 500 events each) 
of proton and Fe induced EAS of vertical incidence for various primary 
energy ranges, $10^{15}$ eV, (1.78-3.16)$\cdot 10^{16}$ eV and 
3.16$\cdot 10^{17}$ eV, calculated with the CORSIKA version 5.62. 
Another set (II) of samples, calculated with a spectral index of the 
power-law slope of $-2.0$ for eight energy ranges (
1$\cdot 10^{16}$ - 1.78$\cdot 10^{16}$ eV; 1.78$\cdot 10^{16}$ - 
3.16$\cdot 10^{16}$ eV; 3.16$\cdot 10^{16}$ - 5.62$\cdot 10^{16}$ eV;
5.62$\cdot 10^{16}$ - 1$\cdot 10^{17}$ eV; 1$\cdot 10^{17}$ - 
1.78$\cdot 10^{17}$ eV; 1.78$\cdot 10^{17}$ - 3.16$\cdot 10^{17}$ eV; 
3.16$\cdot 10^{17}$ - 5.62$\cdot 10^{17}$ eV;
5.62$\cdot 10^{17}$ - 1$\cdot 10^{18}$ eV) with the CORSIKA version 5.64 
and the QGSJET model, comprises smaller number of events (100 or up to 
only 9 for the highest energy range), distributed randomly
over an angle-of-incidence range of $0^{\circ}-30^{\circ}$.

\begin{figure}[t]
\centering
\includegraphics[bb=33 178 567 692,width=14.4cm]{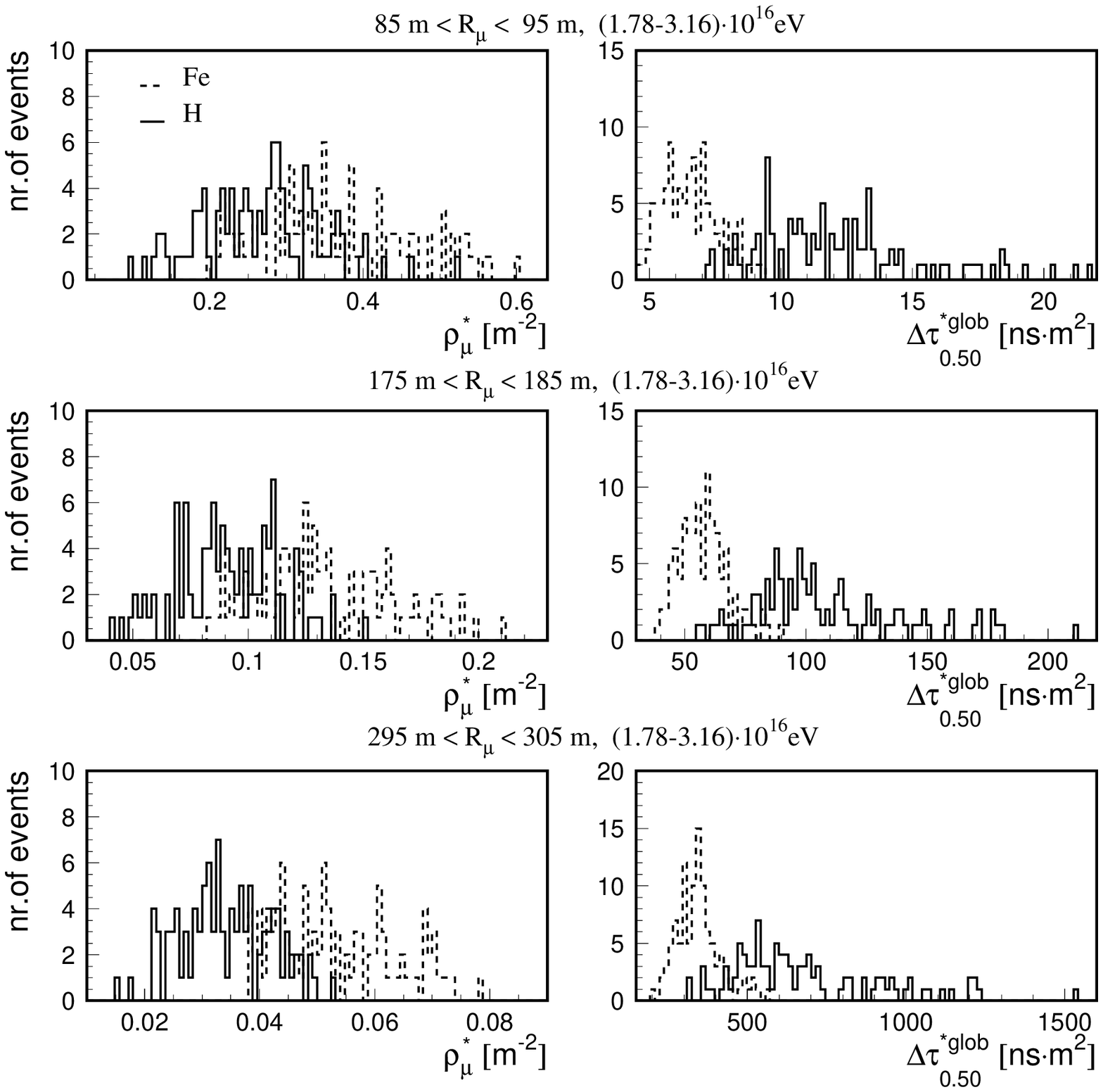}
\caption{\label{Fig. 4} 
Fluctuations of $\rho _\mu ^{*}$ (left)
and the distribution of the reduced time $\Delta \tau _{0.50} ^{*\ glob}$ 
(right) of proton and iron induced EAS ($E_0$ = (1.78-3.16)$\cdot 10^{16}$
eV) at different distances of the shower core.}
\end{figure}

For each simulated EAS event a number of EAS observable have been 
reconstructed: the electromagnetic shower size $N_e$, the
muon content $N_\mu ^{tr}$, the shower age $s$, and in particular 
the time parameters $\Delta \tau _q$ of the muon arrival time 
distributions and their variations with the distance $R_\mu$ from 
the shower axis. For higher primary energies and approximating the
experimental conditions of KASCADE-Grande, instead of $N_e$ the 
quantity $N_{ch}$ (the total number of charged particles) is
preferentially considered. Consequently we deduce also an age 
quantity $s_{ch}$ derived from the lateral distribution of the
charged particles. As approximate energy identifier instead 
of $N_\mu ^{tr}$ the quantity $\rho _\mu$ ($R_\mu$ = 600 m)
with $E_{thr} = 240$ MeV is introduced, which has been often 
considered in the past for this purpose \cite{52,53}.

The muon arrival time distributions have been calculated for muons 
with an energy threshold $E_{thr} = 2.4$ GeV, observed with
a multiplicity $n \ge 3$ per event. The value of this energy threshold is
chosen according to the muon detection threshold of the KASCADE
Central Detector which is foreseen to be used for the time 
measurements \cite{34}. Just for an impression average 
arrival time distributions for proton and iron induced
EAS at two different energies and different muon energy detection 
thresholds are displayed in Fig. 2. The figure indicates
that differences for different kinds of primaries are, if ever, obvious
in the initial part of the distributions,
represented by the first quartile $\Delta \tau_{0.25}$.
It gives also an impression about the order of the needed time resolution
(about $1.5$ ns) to reveal the differences. But generally the present 
paper is not focussed to the discussion of instrumental effects. 

The distributions of the different quantiles $\Delta \tau_{q}$ 
of event-by-event observations have been shown \cite{29,31} to follow 
fairly well a phenomenological parameterisation by $\Gamma$- probability 
distribution functions, with parameters varying with the
primary energy and the distance $R_\mu$.

\begin{figure}[t]
\centering
\includegraphics[bb=38 153 562 677,width=15.0cm]{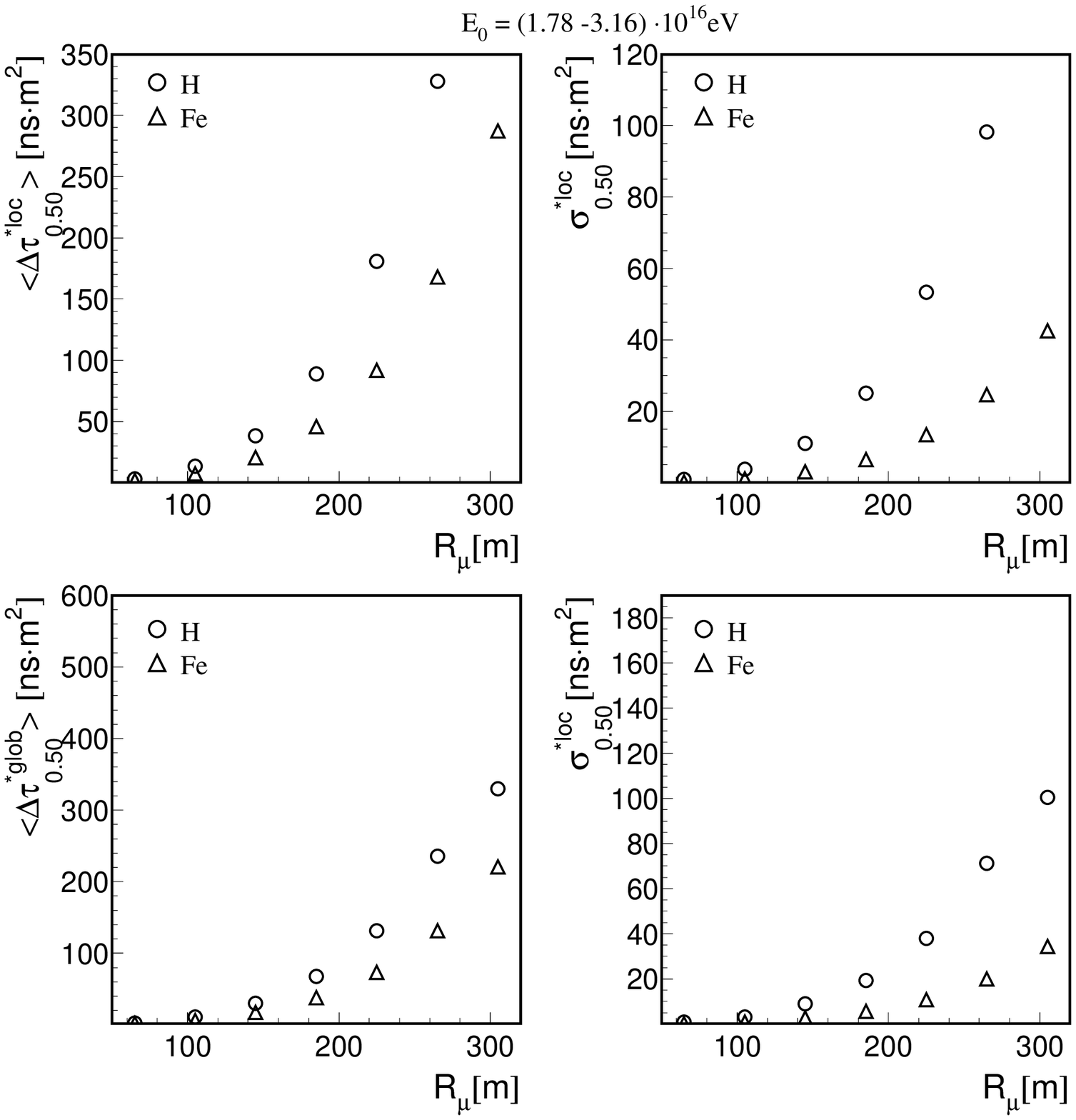}
\caption{\label{Fig. 5} Reduced local and global time profiles 
$<\Delta \tau _{0.50} ^{* \ loc}>$ and $<\Delta \tau _{0.50} ^{*\ glob}>$
of the EAS muon component induced by proton  and Fe primaries of the energy 
range of (1.78-3.16)$\cdot 10^{16}$ eV based on Monte Carlo
simulations using the QGSJET model as generator. }
\end{figure}

The following figures (Figs.~3 and~5) display the time profiles of the 
muon component, i.e. the variation of the mean $<\Delta \tau _q>$ of
the  time parameters and of their fluctuations (characterised  
by the standard deviation $\sigma _q$) with $R_\mu$.
They are calculated with samples of set I and analysed for radial ranges 
up to 300 m (in bins $\Delta R_\mu$ = 10 m).

Fig. 3a compares local and global time profiles $<\Delta \tau _{0.50}>$ 
with respect of differences of the primary mass, while Fig. 3b
displays the time profiles of different quartiles of the proton case. 
Obviously the local and global profiles 
do not exhibit pronounced differences for different primaries. 
The quantities $<\Delta \tau ^{glob}>$ and $<\Delta \tau ^{loc}>$ differ
by an offset $(\Delta \tau ^{glob} _1 )$ which varies with $R_\mu$. 
It has been shown that differences of $<\Delta \tau ^{glob} _1 >$ are in 
the order of 1 ns for different primary muons in the considered range 
of distances from the shower axis \cite{54}. 

Since the lateral distribution of the muon component displays also some 
differences for different  kinds of primaries,
the profile of a combined quantity ("reduced time") 
$\Delta \tau _{0.50} ^{*} = \Delta \tau _{0.50} (R_ \mu)/\rho _\mu ^{*}(R_\mu)$
is of interest where $\rho _\mu ^{*}$ represents the density of
muons with energies $E_\mu > 2.4$ GeV. This quantity is measurable 
with the KASCADE Central Detector and is related to the
multiplicity of muons observed in the EAS event.

As an example in Fig. 4 the fluctuations of $\rho _\mu ^{*}$ are displayed 
for different distances $R _\mu$ from
the shower core, indicating an increasing separation of the distributions 
with increasing $R _\mu$. This feature is
reflected in the distributions of the time quantity 
$\Delta \tau _{0.50} ^{* \ glob}$.

Fig. 5 compares the local $<\Delta \tau _{0.50} ^{* \ loc}>$ and global 
$<\Delta \tau _{0.50} ^{* \ glob}>$ profiles.
A marginally slight improvement of the discrimination of
the global quantity for distances larger than 100 m is indicated. 
This feature gets more pronounced at higher energies.
Though global quantities include the curvature of the shower disc, 
while local quantities display only the internal structure of the EAS. 
At not too large distances from the shower axis the offset between 
global and local profiles is nearly mass-independent.

Generally we note that differences of distributions arising from different 
mass primaries are in the order of
nanoseconds, increasing with the distance from the shower axis. 
For the energies $\le$ (1.78-3.16)$\cdot 10^{16}$ eV (set I) 
also some exploratory calculations, adopting the VENUS
or SYBILL model, have been performed and show that the model 
dependence leads to differences of the distributions
(within the considered $R_\mu$ range) in the order of 2\%
with no clear trend \cite{55}.

\section{Correlated distributions}

The concept of modern cosmic ray experiments like KASCADE or 
KASCADE-Grande with a multi-component detector array is to
deduce the information  from correlated measurements of a larger 
\begin{figure}[!t]
\centering
\includegraphics[bb=10 150 600 660,width=17.0cm]{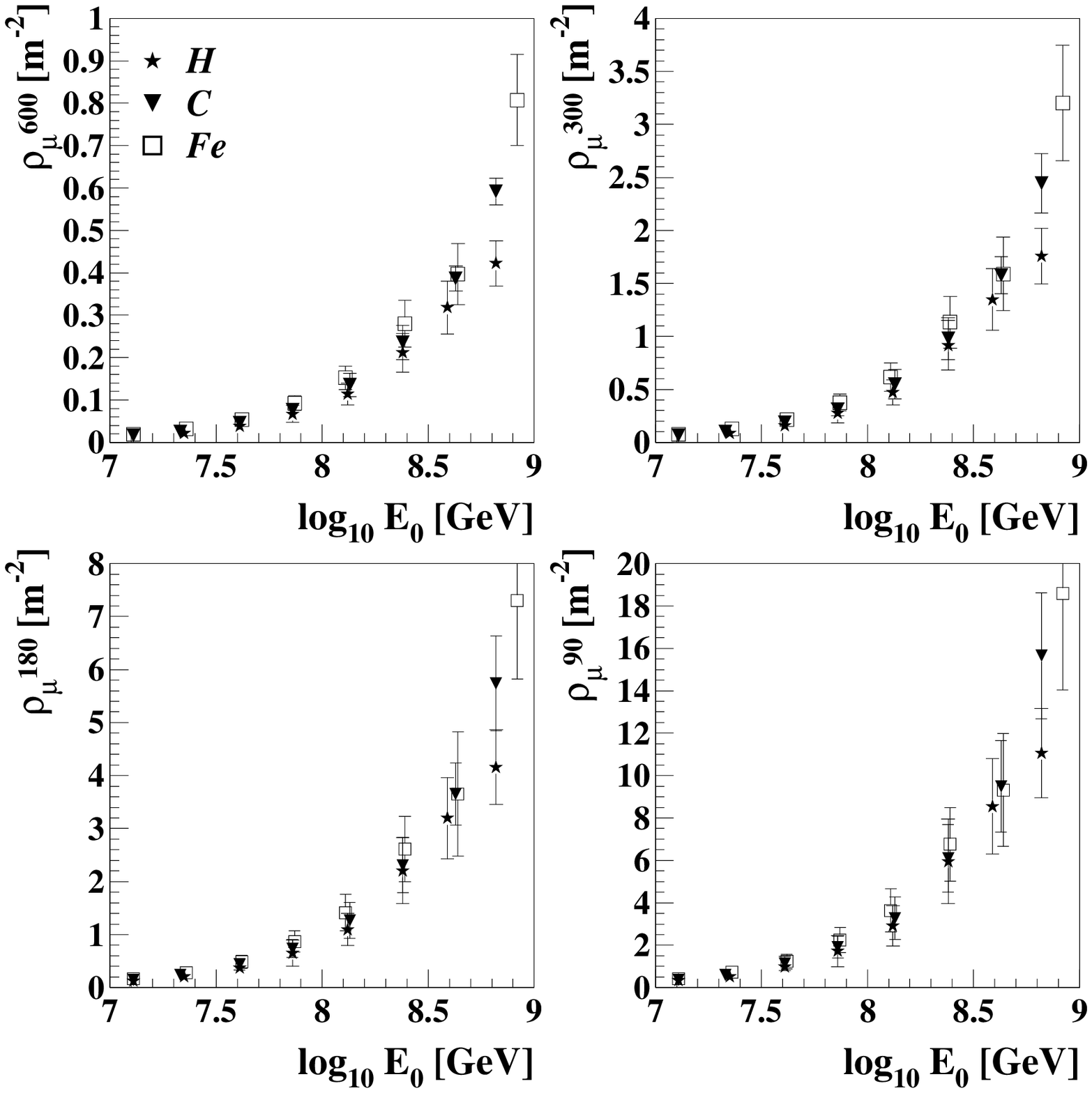}
\caption{\label{Fig. 6}The variation of the primary energy $E_0$ with 
the average of $\rho _\mu (600$ m), $\rho _\mu (300$ m), $\rho _\mu (180$ m)
and $\rho _\mu (90$ m)  for proton, C and Fe induced EAS.}
\end{figure}
number of observable EAS parameters for each individual event
(see Ref. \cite{56}). Particular observable quantities are the 
electromagnetic 
shower size $N_e$, the muon content $N_\mu$ (or the truncated
muon number $N_\mu ^{tr}$ in case of KASCADE) and other specific 
observable quantities, characterising the various EAS components. In view
of the measuring possibilities of the KASCADE-Grande array the 
KASCADE observable $N_e$, $N_\mu$ and $N_\mu ^{tr}$ (the latter loses
also the role as approximate energy identifier for larger energies) are 
not accessible. For KASCADE-Grande they will be replaced by the total
number of the charged particles $N_{ch}$ and a (partial) muon 
number, as measured with the original KASCADE detector array 
being embedded in the KASCADE-Grande installation. This muon 
number is dependent on the distance to the
shower core. In the present studies, for the sake of simplicity, we 
represent it by $\rho _\mu(R_\mu)$ i.e. the muon 
density (with $E_{thr} = 240$ MeV) at $R_\mu$ and consider in particular 
$\rho _\mu (600$ m).

\begin{figure}[!t]
\centering
\includegraphics[bb=30 179 553 687,width=15.0cm]{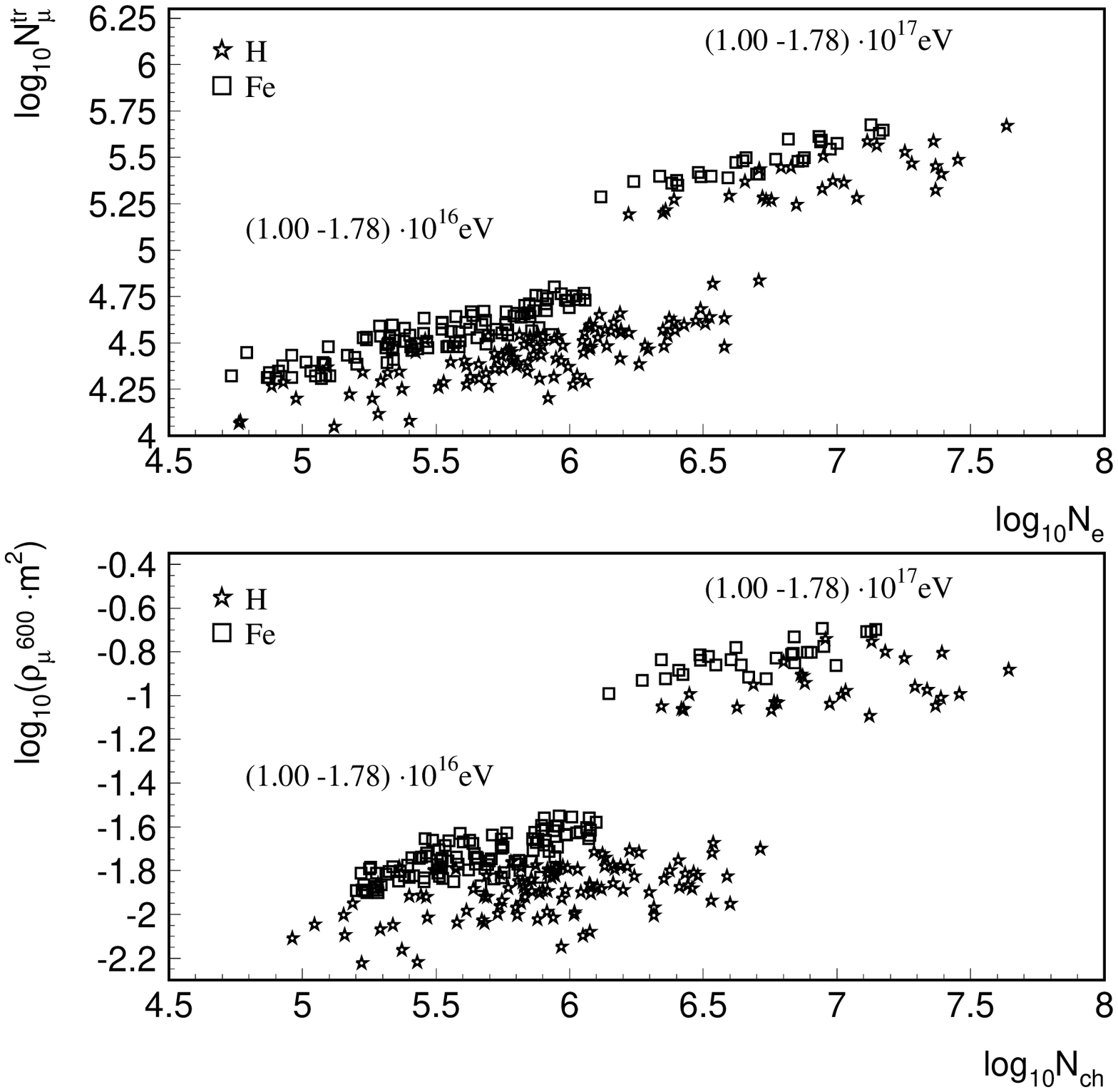}
\caption{\label{Fig. 7} Comparison of the $N_e - N_\mu ^{tr}$ correlation 
with the $N_{ch} -\rho _\mu (600$ m)
correlation at different primary energies.}
\end{figure}

The variation of the muon density $\rho _\mu (R_\mu)$ with the primary 
energy is illustrated in Fig. 6 by results of the present simulations. 
It is obvious 
that $\rho _\mu (600$ m) as well as the muon density at other distances 
(see Fig. 6) do not provide a strictly  mass independent energy
identifier. In addition as known for $N_\mu ^{tr}$, there are 
considerable fluctuations of $\rho _\mu (600$ m) at fixed energies.

The current and foreseen analyses of the experiments attempt to infer 
from the correlations of various observable
an estimate simultaneously for the energy and the mass of the primary 
cosmic particle inducing the EAS event. As
the most powerful correlation the $N_e - N_\mu ^{tr}$ correlation has 
been proven (see Ref. \cite{56}). 
It is expected that the $N_e-\rho _\mu (R_\mu)$ correlation 
will exhibit a similar discrimination power which can be defined in terms 
of the true classification probabilities resulting from the analysis
(see section 4). The features are
indicated in Fig. 7 which compares the $N_e - N_\mu ^{tr}$ correlation 
with the $N_{ch} -\rho _\mu (600$ m) correlation at two different
primary energies.

Correlations with additional observable EAS parameters, though of 
relatively small own classification power, shrink the influence of the 
natural EAS fluctuations and test the consistency of the analysis \cite{57}.

\begin{figure}[!t]
\centering
\includegraphics[bb=33 179 564 692,width=15.0cm]{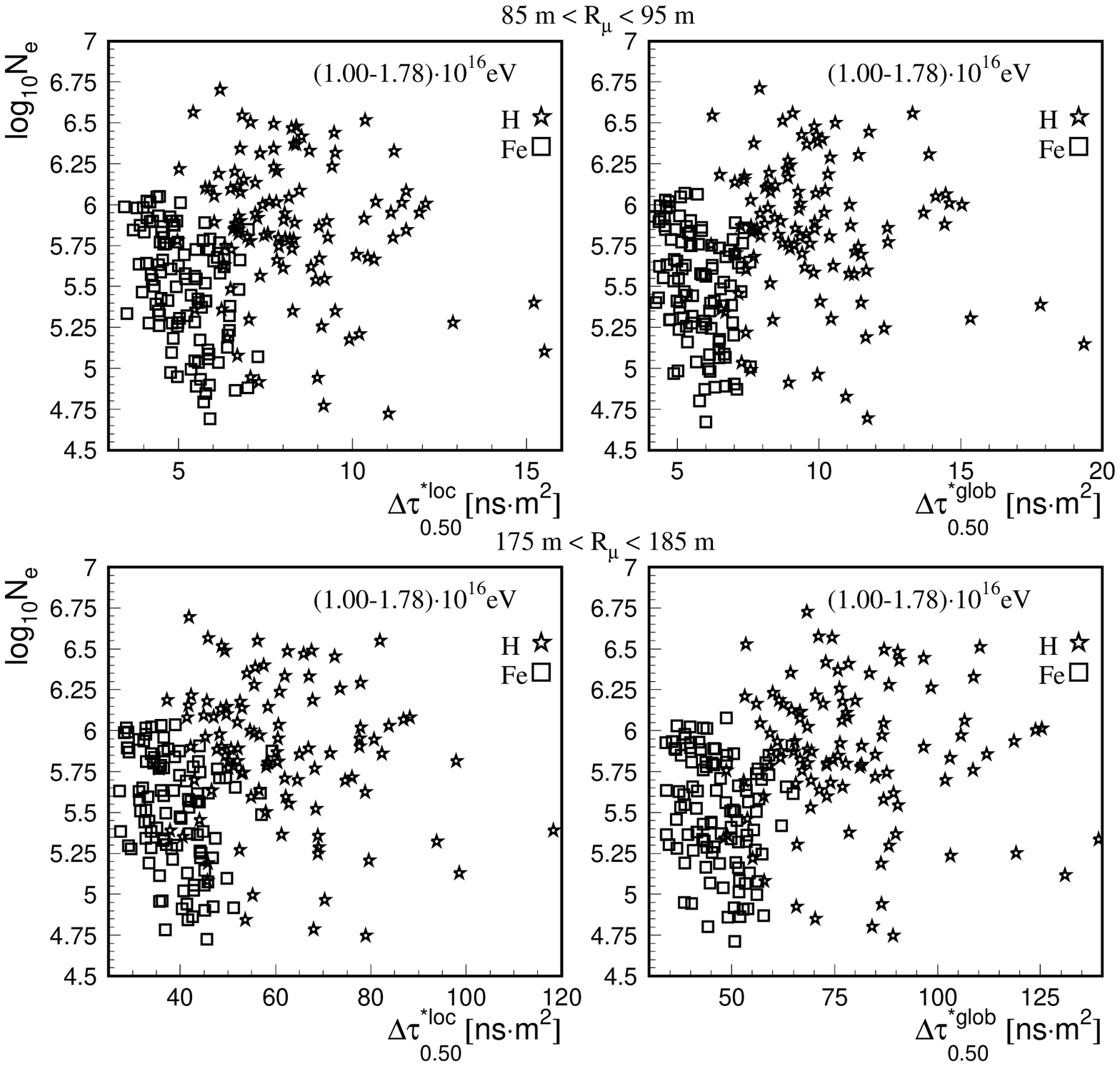}
\caption{\label{Fig. 8}
Correlation of the local quantity 
$\Delta \tau _{0.50} ^{* \ loc} (R_\mu)$ 
and of the global quantity
$\Delta \tau _{0.50} ^{* \ glob} (R_\mu)$ 
with the $log_{10}$ $N_e$
for proton and iron induced showers of the energy  
$E_0 = (1.00-1.78)\cdot 10^{16}$ eV in two different $R_\mu$ ranges.}
\end{figure}

The observation of muon arrival time distributions at particular 
distances $R_\mu$ from the shower centre is combined with the observation
of the local muon density  
$\rho ^* _\mu (R_\mu)$ with $E_{thr} = 2.4$ GeV reflected by the
muon multiplicity i.e. the 
registered number of muons spanning the single
arrival time distribution. It turned out \cite{55} that the 
$\Delta \tau _q (R_\mu) - \rho ^* _\mu (R_\mu)$
correlation improves the mass discrimination and can be fairly well 
replaced by a combined parameter 
$\Delta \tau _q ^* (R_\mu) = \Delta \tau _q (R_\mu)/\rho ^*_\mu (R_\mu)$. 
This result has been already anticipated in the presentation of the 
EAS time profiles in section 2.

Fig. 8 displays the correlation distributions of 
$\Delta \tau _{0.50} ^{* \ loc} (R_\mu)$
and $\Delta \tau _{0.50} ^{* \ glob} (R_\mu)$, respectively,
with the $\log_{10} N_e$ for proton and iron induced showers  
of the energy  $E_0 = (1.78-3.16)\cdot 10^{16}$ eV in two
different $R_\mu$ ranges.

\begin{figure}[t]
\centering
\includegraphics[bb=37 179 564 692,width=14.5cm]{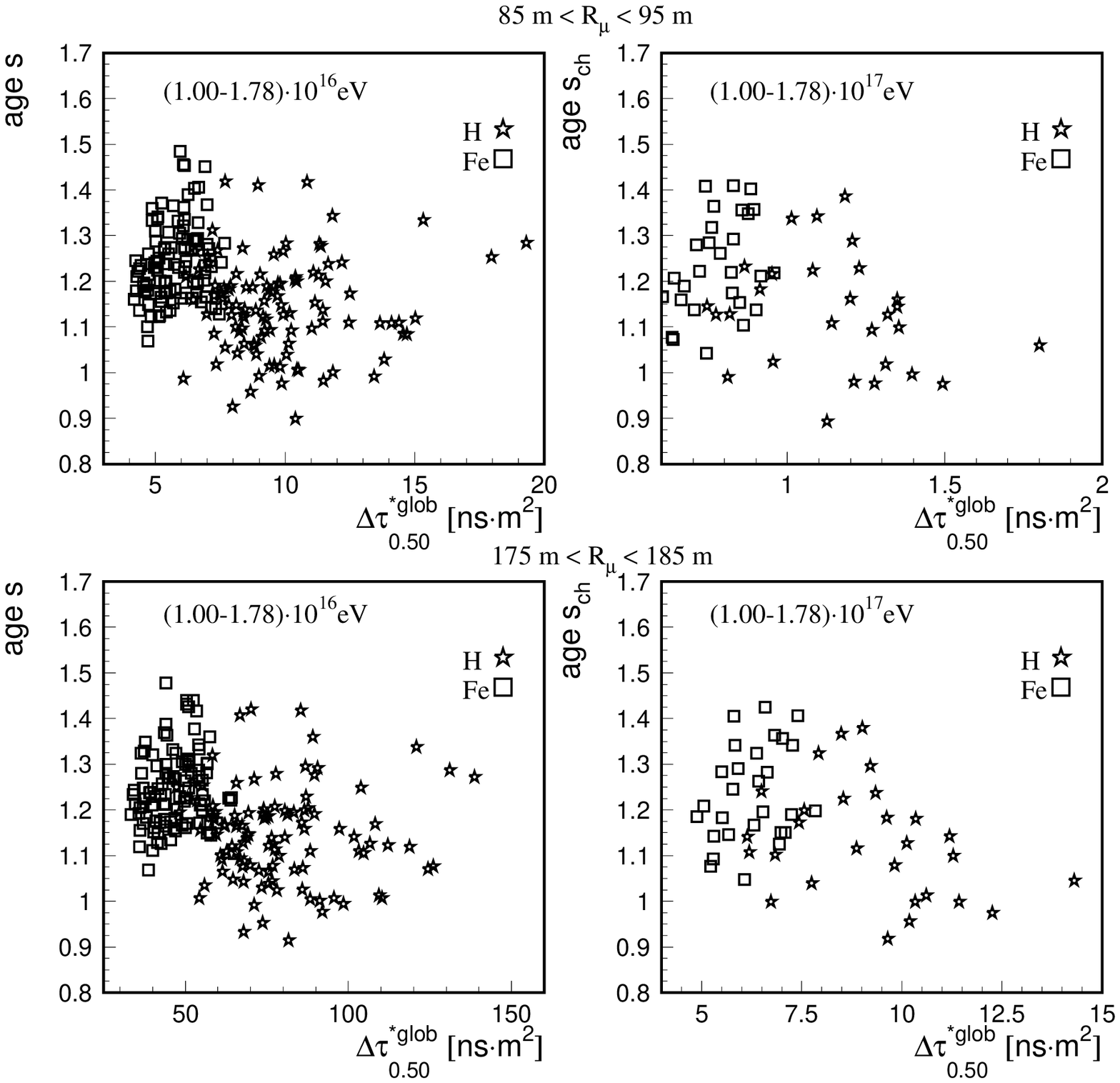}
\caption{\label{Fig. 9}
Correlation of $\Delta \tau _{0.50} ^{* \ glob} (R_\mu)$ with the 
age s and age $s_{ch}$, respectively, 
for proton and iron induced showers of the energy  
$E_0 = (1.78-3.16)\cdot 10^{16}$ eV and 
$E_0 = (1.78-3.16)\cdot 10^{17}$, respectively, 
in two different $R_\mu$ ranges.}
\end{figure}

Special interest arises from the correlation with the so-called 
shower age $s$. This particular EAS
parameter associated with the soft EAS component is supposed to 
carry also some information about the
longitudinal EAS development. In the cascade theory 
(NKG approximation) \cite{49} $s$ is related to the actual
atmospheric depth of the shower development via the lateral 
distribution involving the Moli{\`e}re radius
$R_{moli\grave{e}re}$ (which is a characteristic unit of length 
of the scattering theory) and the ratio $E_0/\epsilon_0$ of the
incident energy to the critical energy. However, the use of the 
age parameter resulting from the simplified
handling of the electromagnetic component by the cascade 
theory does not describe realistically the
electromagnetic component. Rather a full Monte Carlo simulation 
of the electromagnetic component (by use
of the EGS option \cite{51} in the CORSIKA code) calculating the 
total intensity distribution of the electron
component and the lateral electron distribution with all fluctuations, 
has to be required in order to obtain
results being comparable with the reality. In order to derive from 
lateral distributions resulting from the
EGS - Monte Carlo simulation an "age parameter" value (which is 
not defined within the Monte Carlo approach),
the resulting lateral distribution is subsequently fitted by the NKG 
function $f(R/ R_{moli\grave{e}re}$; $N_e$, $s$) in the same
manner as the experimental observations are analysed 
($R_{moli\grave{e}re}$ = 78 m). Thus a (lateral) "age"
is extracted \cite{58}. In a similar way the lateral distributions 
of the charged particles are processed for deducing the age 
parameter $s_{ch}$.
However it should be noted that such a procedure is considered to be only 
a first approximation. The lateral distribution of charged EAS particles
at larger distances from the shower core deviates from the NKG function, 
and an improved parametrisation should be used \cite{59}.
Nevertheless Fig. 9 displays the correlation with the age parameters,
defined by the NKG function. They emphasise the suggestion to exploit in
analyses of experimental data correlations with the age parameters, 
which are obviously related to the longitudinal development and could 
lead to an improved discrimination of protons and iron induced EAS.

In the following section we base such qualitative statements deduced from 
the inspection of the distributions
(shown as examples in Figs. 8 and 9) on quantitative results of 
statistical analyses of multivariate distributions, considering also the 
more complicated case of three mass classes 
(protons: H, carbon: C, iron: Fe).

\section{Non-parametric analyses of the sensitivity to the primary mass}

Non-parametric statistical methods are most efficient and unbiased tools 
for the analysis  of multidimensional observable-distributions in order 
to associate single observed events to different classes (say to different 
masses of the EAS primaries) by comparing the considered events with the
model distributions without using any  pre-chosen parameterisation.
The methods of decision making and the application to cosmic ray data 
analyses are generally described in Refs.\cite{42,56}. They have
been outlined for studies of muon arrival time distributions in 
Refs.\cite{27,29}. The procedures take into account the effects of the 
natural EAS fluctuations in a quite natural way and are able to
specify the uncertainties, by an estimate of the true-classification  
and misclassification probabilities. The classification probabilities 
are determined by the extent to which the likelihood functions of the
single classes, derived from the simulations, are overlapping. Basically 
the results of such pattern recognition methods, using trained neural 
networks or Bayes decision rules, are dependent from the particular
hadronic interaction model generating the reference pattern for the data 
to be studied. In the present analysis of the sensitivity of various 
correlation distributions, prepared from Monte Carlo simulations, the 
so-called one-leave-out test (see Ref. \cite{42}) is applied, 
which determines the probability that a
multidimensional event, taken from  the considered (simulated) 
distribution, will be correctly ("true") or incorrectly 
("false") classified by the procedure.
\begin{figure}[!t]
\centering
\includegraphics[bb=10 160 570 700,width=17.0cm]{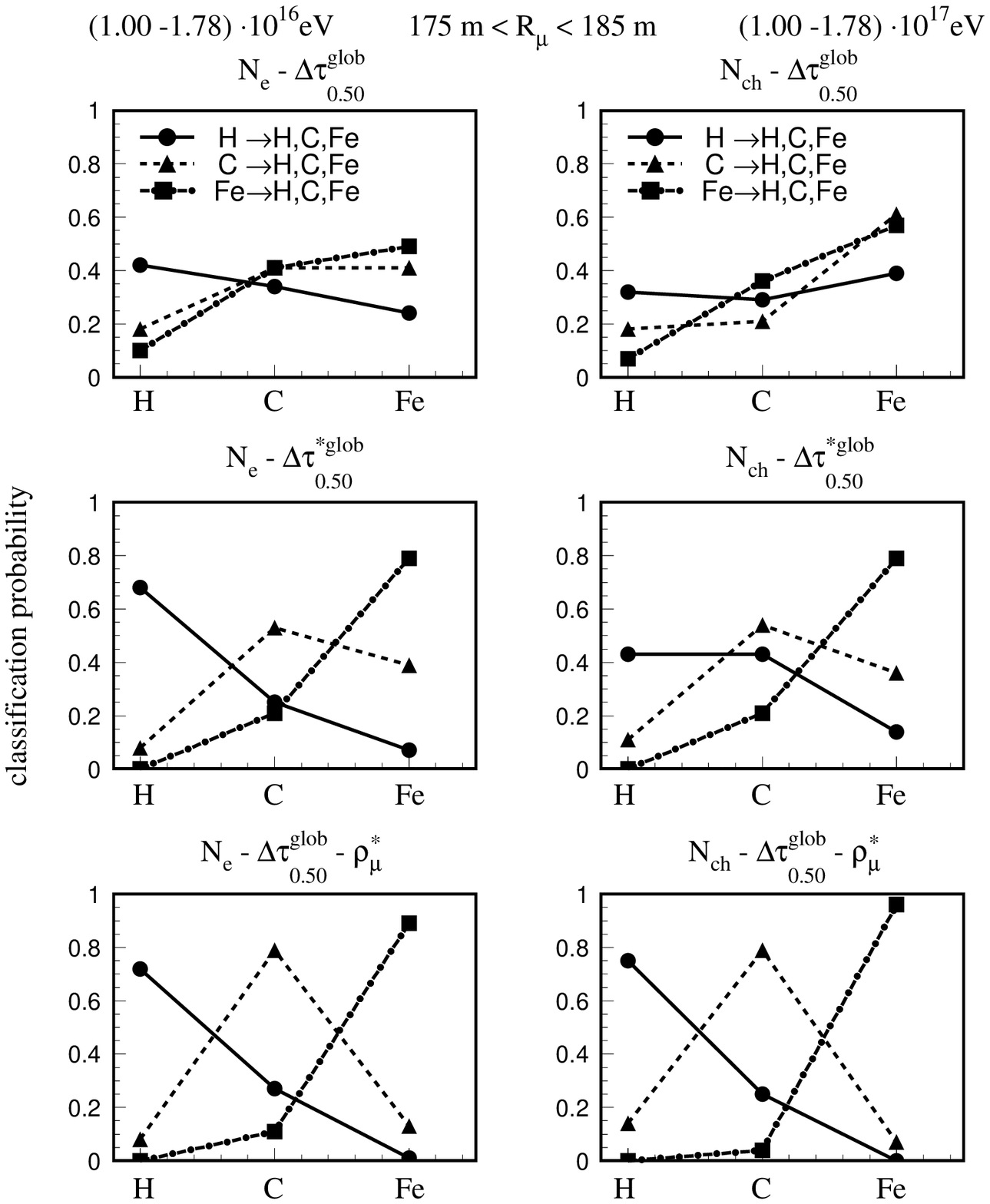}
\caption{\label{Fig. 10} Comparison of the classification probabilities
deduced from the $\Delta \tau _{0.50} ^{\ glob} (R_\mu) - N_e$, 
the $\Delta \tau _{0.50} ^{*\ glob} (R_\mu) - N_e$ and the 
$\Delta \tau _{0.50} ^{\ glob} (R_\mu) - \rho _\mu ^* (R_\mu) - N_e$ 
correlations at the incident energy of (1.00-1.78)$\cdot 10^{16}$ eV 
at $R_\mu $ = 175-185 m (left).
Similarly the classification probabilities 
deduced from the $\Delta \tau _{0.50} ^{\ glob} (R_\mu) - N_{ch}$, 
the $\Delta \tau _{0.50} ^{*\ glob} (R_\mu) - N_{ch}$ and the 
$\Delta \tau _{0.50} ^{\ glob} (R_\mu) - \rho _\mu ^* (R_\mu) - N_{ch}$ 
correlations at the incident energy of (1.00-1.78)$\cdot 10^{17}$ eV 
are shown (right). The lines are drawn for guiding the eyes.}
\end{figure}
\begin{figure}[!t]
\centering
\includegraphics[bb=10 160 580 710,width=17.0cm]{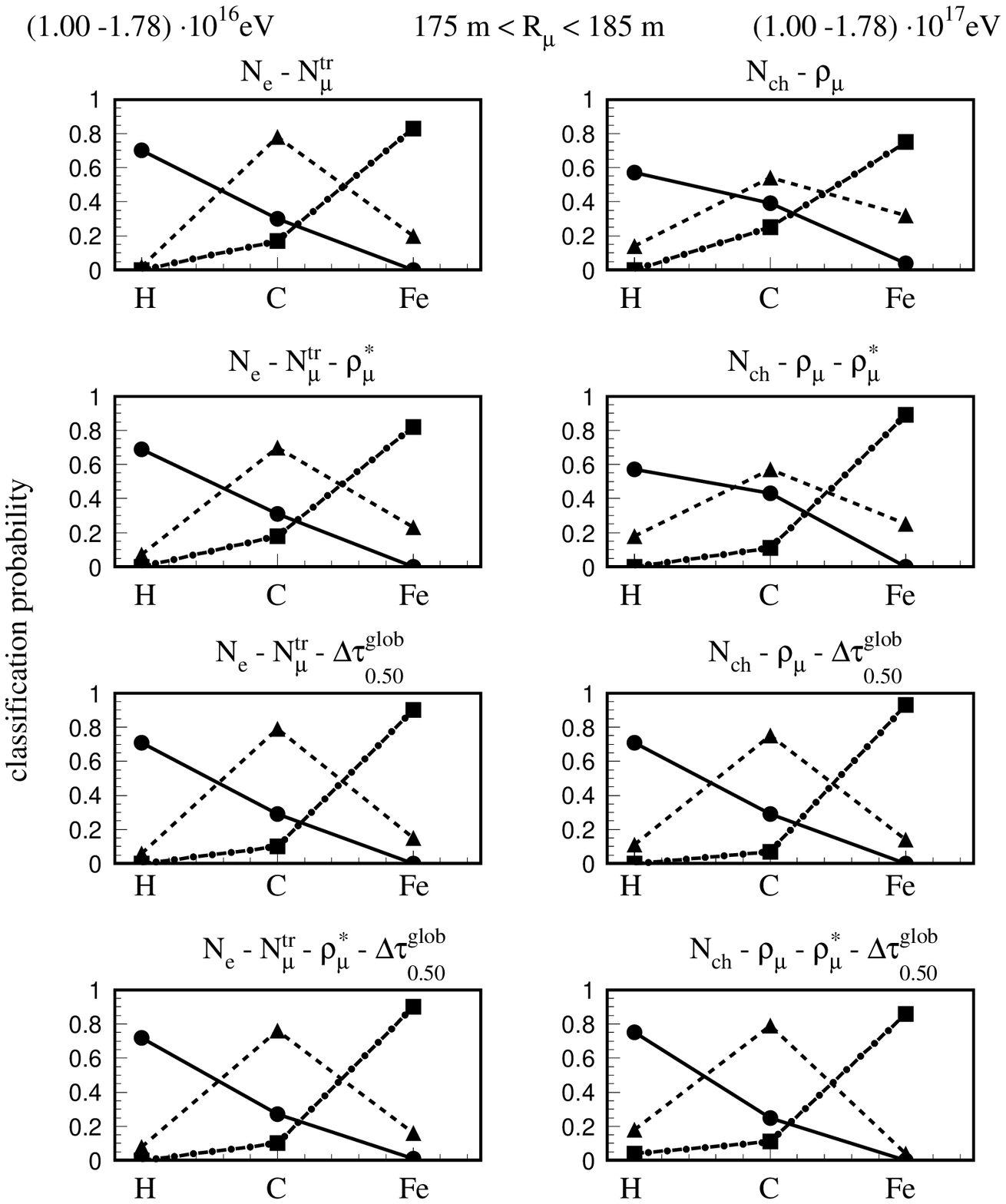}
\caption{\label{Fig. 11} The influence of the global time 
parameter $\Delta \tau _{0.50} ^{\ glob} (R_\mu)$ and
the accompanying $\rho _\mu ^* (R_\mu)$ on the true and
false - classifications of the $N_e - N_\mu ^{tr}$ correlation
shown for the primary energy (1.00-1.78)$\cdot 10^{16}$ eV 
at $R_\mu$ = 175-185 m. Similarly the influence on the true and 
false - classifications of the $N_{ch} - \rho _\mu (R_\mu)$
correlation is shown for the primary energy 
(1.00-1.78)$\cdot 10^{17}$ eV (right). For the explanation of the 
symbols see Fig. 10.}
\end{figure}

\begin{figure}[!t]
\addtocounter{subfigure}{1}
\centering
  \subfigure{
  \includegraphics[bb=50 350 520 530,width=15.0cm]{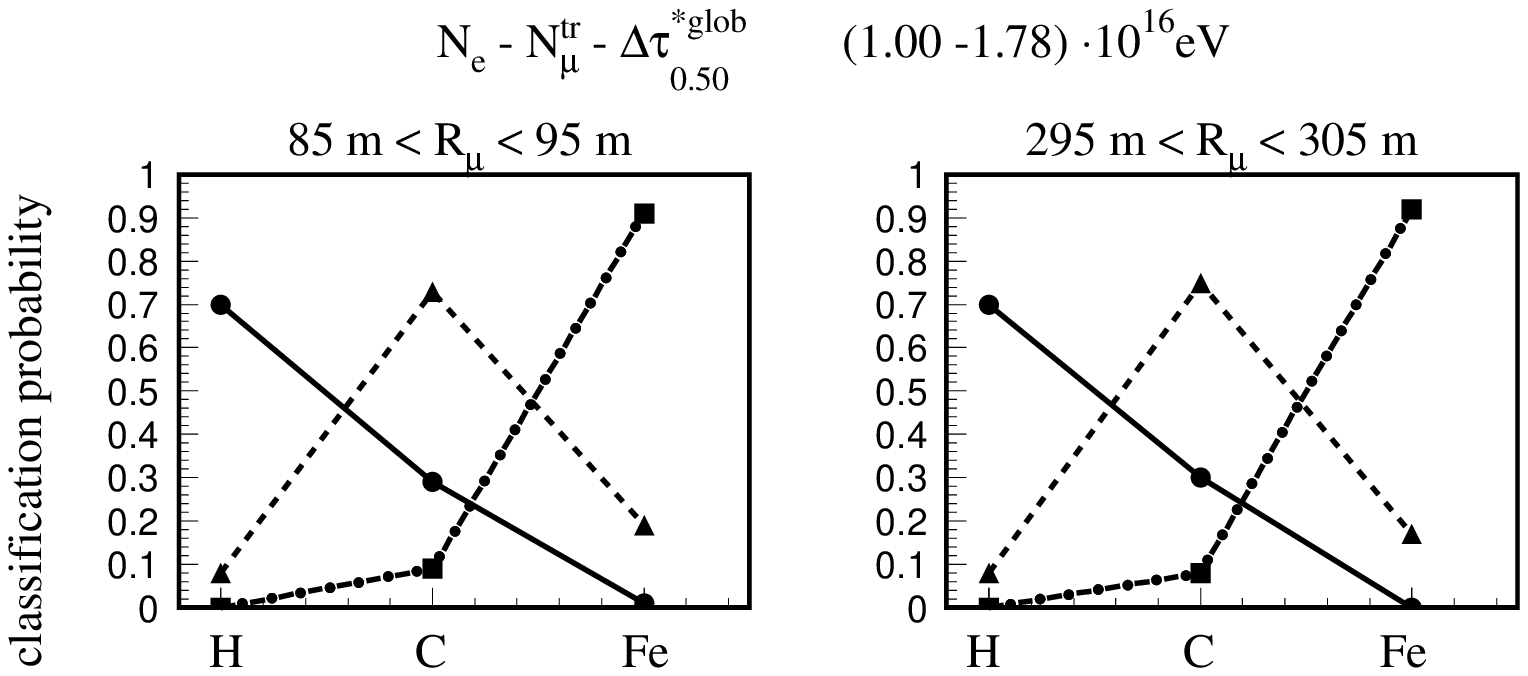}}
\caption{\label{Fig. 12a}{\bf($a$)} The influence of the global time 
parameter $\Delta \tau _{0.50} ^{*\ glob} (R_\mu)$
on the true and false - classifications of the $N_e - N_\mu ^{tr}$ 
correlation shown for two distances $R_\mu$ for the primary energy 
range (1.0-1.78)$\cdot 10^{16}$ eV.
For the explanation of the symbols see Fig. 10.}
\end{figure}
\begin{figure}[!t]
\addtocounter{figure}{-1}
\addtocounter{subfigure}{1}
\centering
  \subfigure{
  \includegraphics[bb=50 260 520 600,width=15.0cm]{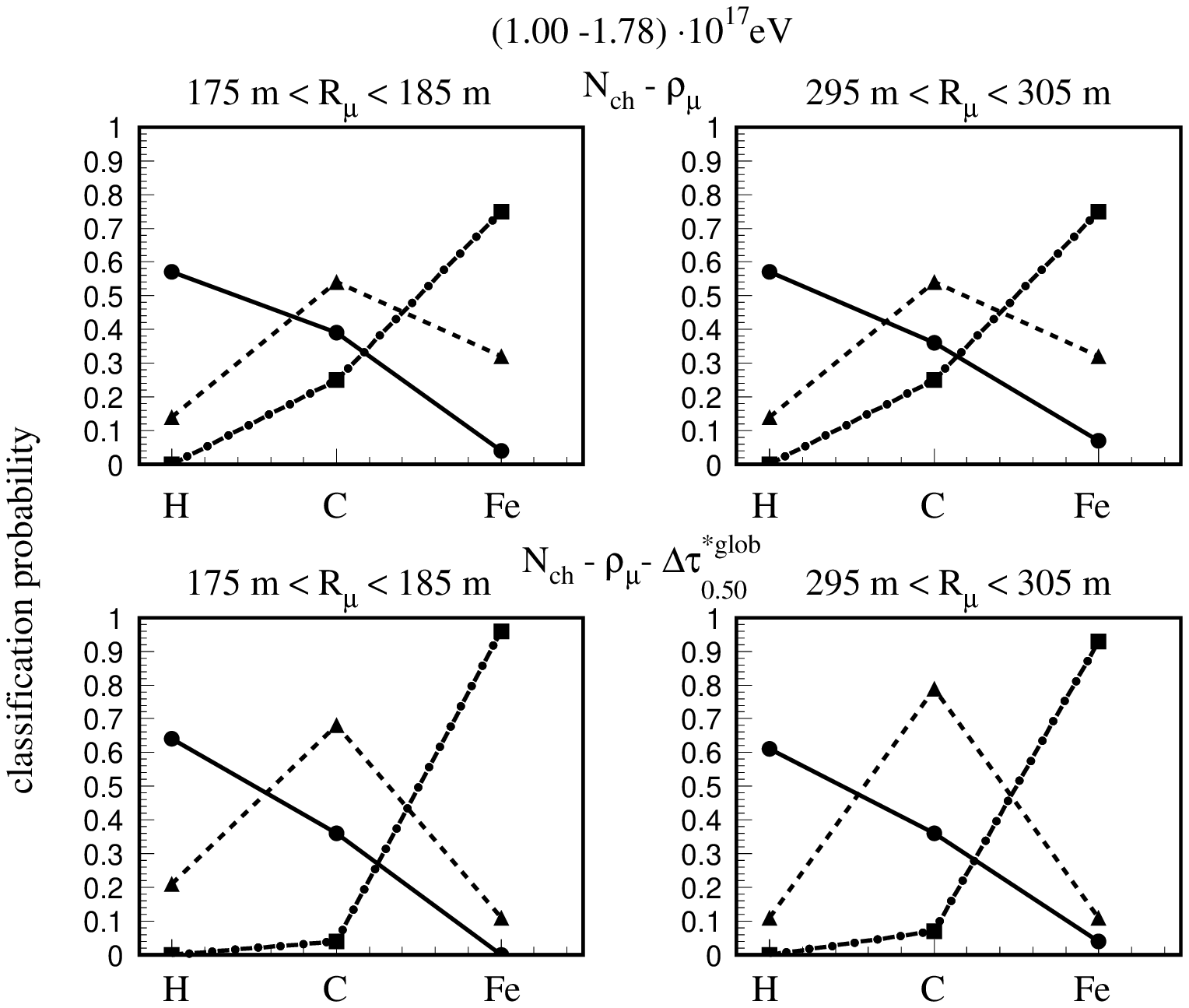}}
\caption{\label{Fig. 12b}{\bf($b$)} The influence of the global time 
parameter $\Delta \tau _{0.50} ^{* \ glob} (R_\mu)$
on the true and false - classifications of the $N_{ch} - \rho _\mu (R_\mu)$ 
correlation shown for two distances $R_\mu$ and for the primary energy 
range (1.0-1.78)$\cdot 10^{17}$ eV.
For the explanation of the symbols see Fig. 10.}
\end{figure}

The following figures present the classification ($H \rightarrow H$; 
$C \rightarrow C$; $Fe \rightarrow Fe$) and misclassification
($H \rightarrow C,Fe$; $C \rightarrow H,Fe$; $Fe\rightarrow H,C$) 
probabilities for various correlations of different EAS observable.
The discrimination power may be quantified by the ``Bayes risk''
defined as
\begin{equation*}
R = \: \frac{1}{M_i} \: \: \sum_{j=1}^{M_i} \varepsilon (x_j)
\end{equation*}
where
\begin{equation*}
\varepsilon (x_j) = \left\{ \begin{array}{ll} 0 & \: \: for \: \: 
 true \: \: classification\\ 1 & \: \: otherwise \end{array} \right.
\end{equation*}
of the event $x_j$ of the sample $M_i$. 
When considering the results based on simulations, one should keep 
in mind that the primary energy $E_0$ is a-priori known, while in the 
real cases $E_0$ has to be simultaneously estimated from the data, in 
particular from the $N_e - N_\mu ^{tr}$ correlation. Within a limited 
energy range $N_\mu ^{tr}$ can be used as approximate energy identifier. 
From this reason  correlations of $N_\mu ^{tr}$ with time observable
are not specially scrutinised.

Even with the primary energy given, the fluctuations of $N_e$ and $N_{ch}$  
or of the correlations $\Delta \tau _{0.50} ^{\ glob} (R_\mu) - N_e$ and 
$\Delta \tau _{0.50} ^{\ glob} (R_\mu) - N_{ch}$ are so large, that the true 
classification probability remains modest. Fig. 10 confirms the 
approximate equivalence of the parameter $\Delta \tau _{0.50} ^* (R_\mu)$
with the $\Delta \tau _{0.50} (R_\mu) - \rho _\mu ^* (R_\mu)$ correlation, 
though the $\Delta \tau _{0.50} (R_\mu) - \rho _\mu ^* (R_\mu)$ correlation 
appears obviously preferable in view of the mass classification. It is
obvious that the quantity $\rho _\mu ^* (R_\mu)$ (simultaneously observed 
with  multiplicity in arrival time measurements of KASCADE (see 
Ref. \cite{34})) leads to a considerable stabilisation of the 
discrimination effect. Especially when observables (like $N_e$ and 
$\Delta \tau _q$) of comparatively weak discrimination power are 
correlated. We conclude that the time quantities $\Delta \tau _q$ should 
be preferentially introduced in combination with $\rho _\mu ^* (R_\mu)$,
if ever experimentally possible.

The effect of global muon arrival time distributions, represented by 
$\Delta \tau _{0.50} ^{\ glob} (R_\mu)$ is quantified by the classification 
probabilities given in Fig. 11. When the detector installation 
is able to measure the $N_e - N_\mu ^{tr}$ correlation, adding 
$\rho _\mu ^* (R_\mu)$ and $\Delta \tau _{0.50} ^{\ glob} (R_\mu)$ has only 
a minor discrimination effect, even at larger distances from the shower 
core. This is in contrast to the case, when only $N_{ch}$  and a partial 
muon number (in our analysis approximated by $\rho _\mu (R_\mu)$)
are experimentally accessible.
At larger distances, especially for the higher primary 
energies, the correlation of muon arrival times has a clear effect of 
improving the mass discrimination. It is interesting to note that 
this result holds also approximately for local time parameters.

Fig. 12a displays the $R_\mu$-dependence of the influence of the global time 
parameter $\Delta \tau _{0.50} ^{* \ glob} (R_\mu)$ on the true and 
false - classifications of the $N_e - N_\mu ^{tr}$ correlation  for the 
primary energy range (1.0-1.78)$\cdot 10^{16}$ eV. At a first glance 
surprisingly, there is no significant tendency with increasing $R_\mu$. 
This feature can be understood with the above arguments of the minor 
contribution of the muon arrival time parameter when the ($R_\mu$ 
independent) $N_e - N_\mu ^{tr}$ correlation can be determined.

However, due to the generally weaker discrimination power of the 
$N_{ch} - \rho _\mu (R_\mu)$ correlation the influence of 
$\Delta \tau _{0.50} ^{\ glob} (R_\mu)$ appears significant in this case 
(Fig. 12b), though also not significantly varying with the distance.
For example the improvement in terms of the Bayes risk for the case of
$E_0 = 10^{16} eV$ is only 2\%-3\% while in the case of $E_0 = 10^{17} eV$
it is 14\%-15\%, i.e. not negligible.

\begin{figure}[t]
\centering
\includegraphics[bb=10 240 540 640,width=16.0cm]{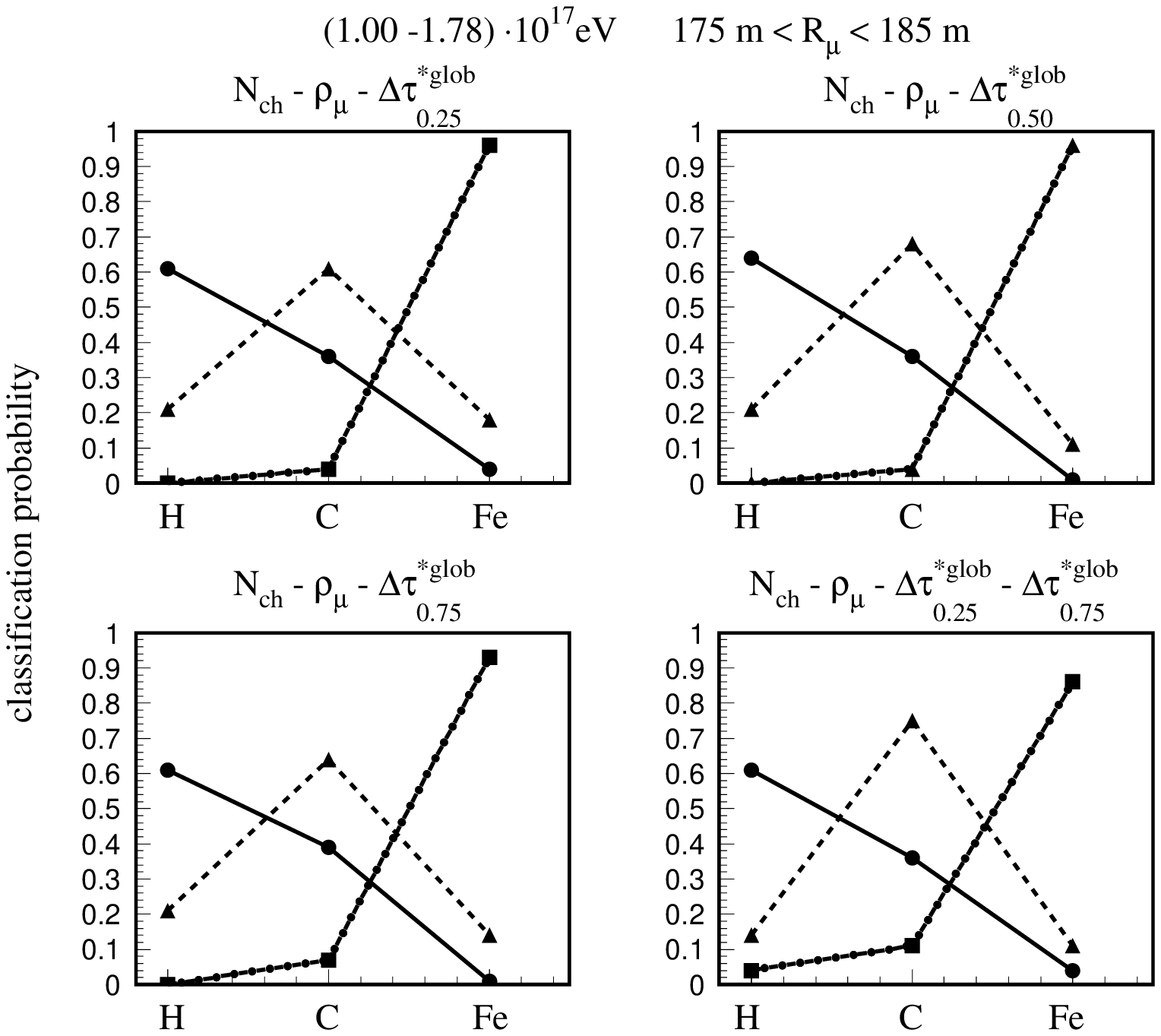}
\caption{\label{Fig. 13} Comparison of the influence of different
quartiles of global arrival time distributions on the 
$N_{ch} - \rho _\mu$ correlation.
For the explanation of the symbols see Fig.~10.}
\end{figure}

As indicated in Fig. 13 differences in the discrimination effects 
of different quartiles 
$\Delta \tau _q$ are hardly obvious and may be obscured by the 
inherent fluctuations.
Studies of the classification and misclassification probabilities of 
EAS observable correlations including the age parameters show a
non-negligible sensitivity of these EAS parameters. However, when 
alone correlated with $N_e (N_{ch})$ such a correlation (say $N_e - s$)
proves to be a very uncertain and instable discriminator due to the fact
that fluctuations of both related observables do obviously corroborate.
Adding an observable of another EAS component stabilises the 
classification result. 

From the classification and misclassification studies 
following features and tendencies can be recognised:

\begin{itemize}
\item The reduced time parameter 
$\Delta \tau _q ^* = \Delta \tau _q (R _\mu)/\rho _\mu ^* (R _\mu)$
(where $\rho _\mu ^*$ represents the density of muons with energies 
$E _\mu > 2.4$ GeV)
absorbs partially the $\Delta \tau _q (R _\mu) - \rho _\mu ^* (R _\mu)$
correlation and enhances the mass classification sensitivity.

\item The most important features are condensed in Figs. 11 and 12, 
revealing a
significant contribution of the muon arrival time information when 
correlated with the EAS observable $N_{ch}$ and $\rho _\mu (R _\mu)$, 
experimentally accessible by the foreseen layout of KASCADE-Grande.

\item The shower age parameters $s$ and $s_{ch}$ (which have been scrutinised
as additional parameters) have modest, but
non-negligible effects for the mass discrimination.

\item Correlations of largely fluctuating quantities, 
resulting from related observations (like $N_e$ and $s$, e.g.), lead to a bad 
classification with large uncertainties, but adding the correlation with
an observable from another EAS component (like $\Delta \tau _q ^*$), even 
also fluctuating, induces an considerable improvement.
\end{itemize}

As already mentioned - as additional calculations show - the model 
dependence of the results of the classification probabilities is 
rather marginal. 
However, it should be remarked, this feature could be in detail different for 
different EAS parameter correlations and is also not yet explored 
for higher energy ranges.

\section{The efficiency of the observation conditions and 
the energy and mass dependence}

\begin{figure}[!t]
\centering
\includegraphics[bb=32 154 562 675,width=15.0cm]{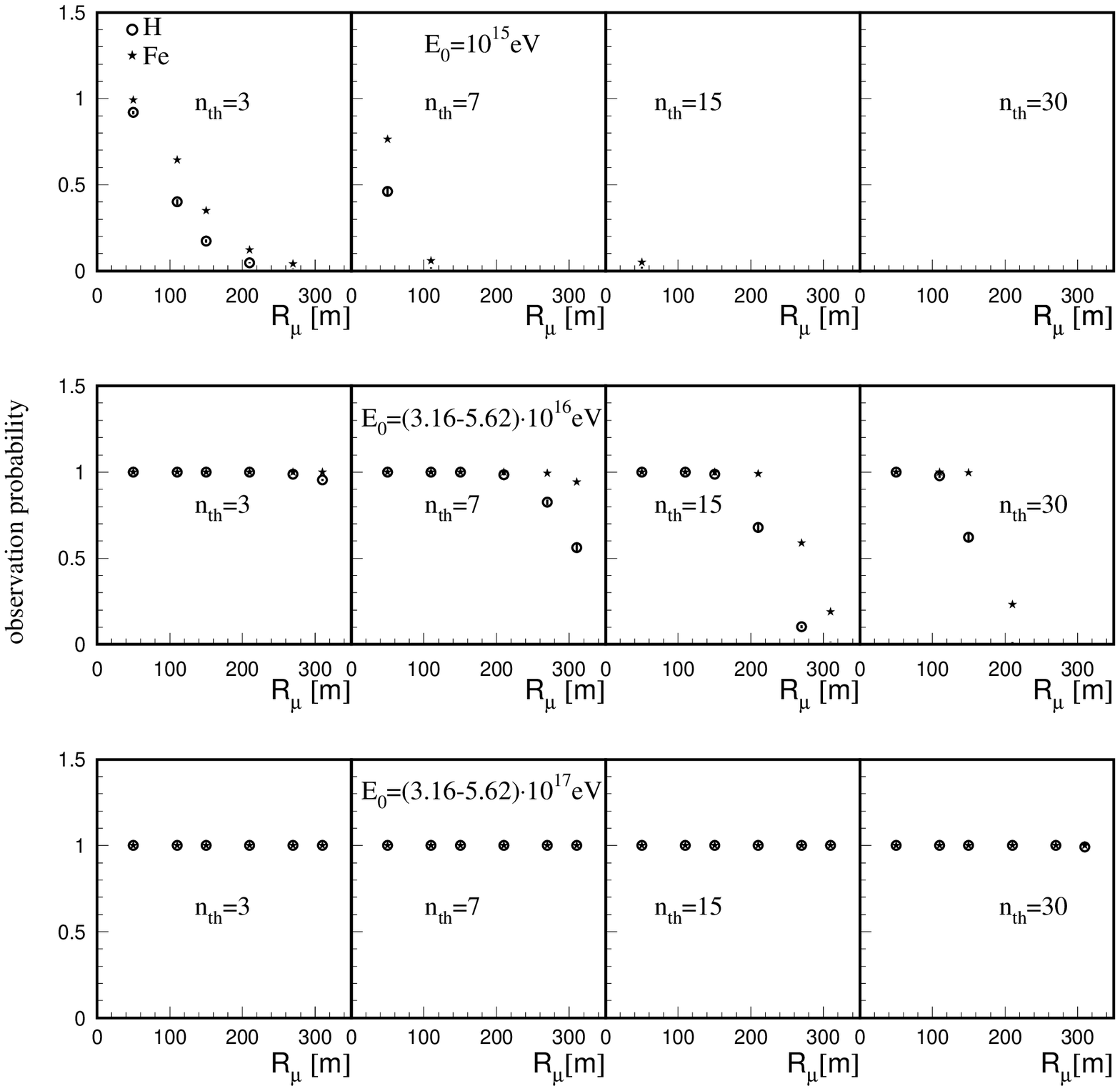}
\caption{\label{Fig. 14} The probabilities to register EAS events  
from proton and Fe induced EAS of different primary energies under the 
observation conditions ($E_{thr}$ =2.4 GeV, $n_{thr}$, $R _\mu$) with
a $10 \cdot 10$ m$^2$ ideal detector. The variation with $R _\mu$ and 
with the multiplicity threshold $n_{thr}$ is displayed.}
\end{figure}

By the observations of muon arrival time distributions special samples 
of all observed EAS events are selected due to the particular 
observation conditions, 
especially by the energy threshold $E_{thr}$ (= 2.4 GeV in the actual case) 
of the detected muons and the multiplicity threshold $n_{thr}$, i.e. the
number of muons necessary to define an event with an observable arrival 
time distribution. Thus the event selections are affected by the lateral 
distribution and the energy spectrum of EAS muon. Consequently the
probability being accepted in the event sample depends on the observation 
distance $R _\mu$, on $n_{thr}$ and on the mass of the EAS primary. As 
discussed in Ref. \cite{34} this feature implies in experimental
observations of EAS samples efficiency corrections \cite{60} and enables 
consistency tests by varying $R _\mu$ and $n_{thr}$
in the measurements.

Fig. 14 displays the results of calculations of the efficiency to 
register an EAS event under the specified conditions of muon arrival 
time measurements.
 
The efficiency depends strongly on $R _\mu$ and $n_{thr}$ at lower primary
energies ($10^{15}$ eV) in a qualitatively understandable way. The dependence
gets weaker at higher primary energies. In general the dependence and the 
variation with $R _\mu$ and $n_{thr}$ are different for different primary 
masses.

\section{Conclusions}

In the present paper based on realistic EAS Monte Carlo simulations the 
role of muon arrival time distributions has been scrutinised in view of 
correlations with the main EAS observables to be measured with the
KASCADE-Grande layout for the discrimination of the primary mass at higher 
primary energies. The main observables for that purpose are the total 
number of the charged particles $N_{ch}$ and a part of the EAS muon 
intensity $N_\mu ^{part}(R_\mu)$ registered with the KASCADE array, 
embedded in KASCADE-Grande. The partial muon number $N_\mu ^{part}(R_\mu)$ 
is dependent from  the location of the EAS centre. For the present 
simulation studies it has been replaced by the density $\rho _\mu (R_\mu)$ 
($E_\mu > 240$ MeV). In summary, the analyses of the classification and 
misclassification probabilities, determined with non-parametric statistical 
techniques give evidence that the information from muon arrival time 
distributions has a potential for improving the mass
discrimination. This conclusion, being in contrast to findings at lower 
energies, holds for global time quantities as well as for the local ones. 
Actually the EAS profiles of the two different types of time parameters 
differ by an offset being only marginally dependent from the mass. 
This is due to an increasing ``flatness'' of the $<\Delta \tau _1>$
profile with increasing primary energy, and differences are moved to 
larger distances out of experimental reach.
This feature is important with respect to the experimental difficulties
to define $\tau _c$ and to measure global time quantities. 
Obviously in the considered $R_\mu$ range the local times provide
most of the accessible information.

The measuring techniques of muon arrival time distributions 
with the Central Detector of KASCADE imply the
simultaneous determination of the multiplicity $n$ of muons
spanning the arrival distributions. This multiplicity is
related to the density $\rho_\mu ^* (R _\mu)$ of the higher 
energy muons detected with the facilities of the Central Detector.
The quantity $\rho _\mu ^* (R _\mu )$ (or the  number of muons 
experimentally detected with the timing facility),
when correlated with timing measurements (or used as 
$\Delta \tau _q ^* = \Delta \tau _q (R _\mu /\rho _\mu ^* (R_\mu)$)
plays an important role in improving the effects of the arrival 
time observations.

Additionally (as outlined in Ref.\cite{34}) as consequence of the 
specific observation conditions muon arrival
time observations establish a selected subset of all showers, 
and the determination of the mass composition needs
a corresponding correction \cite{60}. Thus the results inferred with a 
variation of the multiplicity threshold $n_{thr}$
and of the observation distance $R_\mu$ provide a consistency test for the 
model used for the non-parametric analyses of the observed EAS.

It should be noted that the present results refer to ideal cases 
since any detector response functions, detector
efficiencies and  limitations of the statistical accuracies due 
to limited detector sizes  have been ignored.
Additionally the necessarily limited number of simulated EAS 
implies a limit on the reasonable number of observables used 
in a single analyses. 

\ack

The Deutsche Forschungsgemeinschaft and the Centre of Excellence 
IDRANAP in the National Institute for Physics 
and Nuclear Engineering, Bucharest have considerably supported 
the present studies. Some of us (I. M. B., H. R. and C. A.) would 
like to thank Prof. Dr. H. Bl{\"u}mer and Prof. Dr. D. Poenaru for 
the kind hospitality in Forschungszentrum Karlsruhe and in National 
Institute for Physics and Nuclear Engineering - Horia-Hulubei, 
Bucharest, respectively, experienced during various mutual research 
visits related to the reported studies.

\end{document}